\documentclass[sn-nature]{sn-jnl}
\usepackage{graphicx}
\usepackage{amsmath,amssymb}
\usepackage{siunitx}
\usepackage{bm}
\usepackage{placeins}
\usepackage{float}
\usepackage{upgreek}
\usepackage{geometry}
\title{Enhancement of far-field thermal emission via polaritonic cavity modes}
 
\author[1,2]{\fnm{Maelie} \sur{Coral}}
\author[3]{\fnm{Jose} \sur{Ordonez-Miranda}}
\author[1,2,4]{\fnm{Georges} \sur{Hamaoui}}
\author[1,2]{\fnm{Roman} \sur{Anufriev}}
\author[1,2]{\fnm{Laurent} \sur{Jalabert}}
\author[1]{\fnm{Masahiro} \sur{Nomura}}
\author[5]{\fnm{Yanick} \sur{De Wilde}}
\author[1,2]{\fnm{Sebastian} \sur{Volz}}

\affil[1]{\orgname{Institute of Industrial Science, The University of Tokyo}, \orgaddress{\city{Tokyo}, \postcode{153-8505}, \country{Japan}}}
\affil[2]{\orgname{LIMMS, CNRS-IIS IRL 2820, The University of Tokyo}, \orgaddress{\city{Tokyo}, \postcode{153-8505}, \country{Japan}}}
\affil[3]{\orgname{Sorbonne Universit\'e, CNRS, Institut des Nanosciences de Paris (INSP)}, \orgaddress{\city{Paris}, \postcode{75005}, \country{France}}}
\affil[4]{\orgname{ESYCOM lab, UMR 9007 CNRS, Univ Gustave Eiffel}, \orgaddress{\city{Marne-la-Vall\'ee}, \postcode{77454}, \country{France}}}
\affil[5]{\orgname{Institut Langevin, ESPCI Paris, Universit\'e PSL, CNRS}, \orgaddress{\city{Paris}, \postcode{75005}, \country{France}}}
\date{} 

\abstract{
Controlling thermal emission is crucial for applications involving thermophotovoltaics, thermal sensing, imaging, and camouflage. While prior studies focused on the emission of thermally excited guided modes (TEGMs) inside cavities, their contribution to the far-field radiation outside cavities has remained unexplored. Here, we demonstrate a tunable far-field thermal channel enabled by TEGMs arising from the coupling of surface phonon-polaritons and cavity resonances. By combining infrared emissivity experiments with fluctuational electrodynamics simulations, we identify distinct spectral features marking the conversion of two-dimensionally confined polaritonic modes into three-dimensional radiative channels. We find that silicon cavities covered with SiO$_2$ enhance the emissivity by up to $200\%$ near the polaritonic spectral resonance, whereas bare silicon cavities yield only broadband enhancement. These findings provide experimental evidence of TEGMs and establish a simple cavity architecture as an effective and scalable platform for tailoring thermal radiation without complex nanofabrication.}


\begin{document}

\maketitle

Thermal radiation is a universal mechanism for energy exchange between matter and its environment, governed by the spectral and angular emissivity of surfaces. In thermodynamic equilibrium, Kirchhoff’s law establishes that emissivity equals absorptivity at each wavelength and direction, directly linking thermal emission to an optical response \cite{Planck14,Kifchhoff60}. Controlling this radiative channel is central to technologies such as thermophotovoltaics, passive radiative cooling, infrared sensing, and thermal camouflage \cite{Modest2013,Basu2009,Chen2020,Lee2023,Greffet02,Cuevas18,Said24}. In particular, the ability to tailor the emissivity within specific spectral bands, such as the mid-infrared atmospheric window ($8-13~\upmu$m), has enabled cooling powers exceeding $100~$Wm$^{-2}$ under direct sunlight \cite{Raman14} and enhanced thermophotovoltaic efficiencies beyond conventional limits \cite{Lenert14}. Recent advances in nanophotonics demonstrated that thermal emission can be engineered far beyond blackbody limits through the excitation of confined electromagnetic modes, including surface phonon-polaritons (SPhPs), plasmonic resonances, and photonic cavity modes \cite{Schuller10,Liu11,Basu2009,Costantini15,Wu20}. These modes dramatically enhance the local density of states (LDOS), leading to strong spectral and spatial confinement of thermal energy. In particular, SPhPs supported by polar dielectrics such as SiO$_2$ or SiC, exhibit resonant behavior in the mid-infrared and can give rise to enhanced and spectrally selective thermal emission, when coupled to structured materials \cite{Caldwell14,Greffet97,Liu11,Landy08,Sathwik24,Xu21}. Parallel efforts explored nanoscale surface patterning, photonic crystals, and metamaterials to achieve precise control of emission, often at the cost of complex nanofabrication and limited scalability \cite{Costantini2019,Li2011,Chen24,Ren24}.

\begin{figure}[H]
\centering
\includegraphics[width=0.9\linewidth]{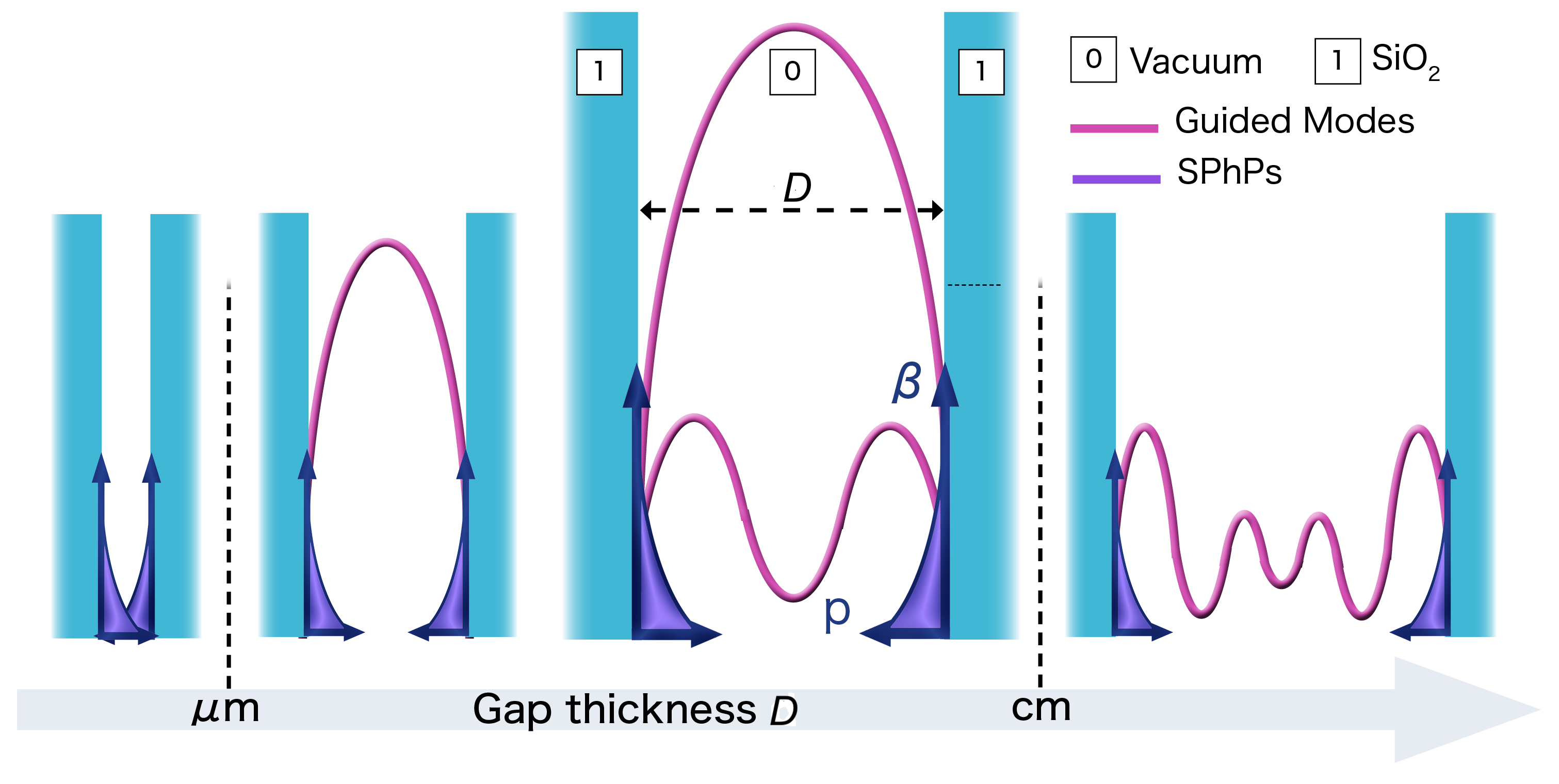}
\caption{\textbf{Hybridization of SPhPs with guided modes inside a planar cavity}. Schematic illustration of the polaritonic cavity concept considered in this work. The planar cavity is formed by two parallel interfaces supporting hybrid electromagnetic modes arising from the coupling between SPhPs (purple zones) and guided cavity modes (pink lines) propagating with in-plane and cross-plane wavevectors $\beta$ and $p$, respectively. The relative balance between near-field coupling, guided propagation, and mode suppression depends on the cavity gap thickness $D$: sub-micrometer gaps ($D\lesssim 1~\upmu$m) favor strongly coupled interface modes, micrometer gaps ($D\gtrsim 1~\upmu$m) enable guided polaritonic propagation, and large gaps ($D\sim 1~$cm) suppress the cavity contribution through absorption. This scheme is based on previous theoretical predictions~\cite{Volz2022} and we use it here to motivate the emergence of thermally excited guided modes (TEGMs) in a vacuum cavity with SiO$_2$ walls.}
\label{fig:fig1}
\end{figure}

The near-field radiation and intracavity energy densities have been widely studied in terms of electromagnetic modes confined along material interfaces. While fluctuational electrodynamics provides a robust framework to describe these phenomena \cite{Rytov53,Rodriguez2013}, the mechanisms by which such confined modes contribute to far-field thermal radiation remain insufficiently understood. The conversion of laterally confined surface modes into propagating radiative channels requires coupling mechanisms that break translational symmetry, such as surface roughness, patterning, or cavity geometries. Although directional and coherent thermal emission was demonstrated in periodic structures and gratings \cite{Sai01}, the role of simple and scalable geometries on this near-to-far-field conversion remains largely unexplored \cite{Baranov19}. Planar cavities formed by parallel interfaces offer a particularly attractive platform to investigate this conversion. In contrast to a single interface that can host only one confined mode, the two parallel interfaces of a planar cavity enable the propagation of multiple hybrid modes due to their mutual coupling \cite{Volz2022,Francoeur08,Tachikawa2024}. The resulting dispersion relation with multiple branches lies below the vacuum light line, revealing the evanescent nature of the modes propagating along the cavity interfaces, while being coupled across the gap. The existence and propagation of these thermally excited guided modes (TEGMs) was theoretically predicted and experimentally observed in planar cavities \cite{Volz2022,Tachikawa2024}. In these structures, the cavity gap thickness $D$ tunes the emergence of TEGMs through the SPhP-waveguide coupling, as shown in Fig.~\ref{fig:fig1}. For sub‑micrometer gaps ($D\lesssim 1~\upmu$m), the radiative flux is dominated by near-field coupling of SPhPs appearing at opposite cavity interfaces, whereas for micrometer‑scale gaps ($D\gtrsim 1~\upmu$m), guided modes progressively emerge and propagate along the in-plane directions, while remaining confined in the cross-plane direction. This spatial confinement enhances thermal emission through an increased number of available modes within the cavity \cite{Volz2022}. For gaps beyond $\sim$1~cm, the material absorption suppresses the guided propagation and effectively quenches the TEGM contribution. 

Despite this understanding on TEGMs, their contribution to the far‑field thermal radiation outside cavities is not rigorously quantified yet. Prior studies mainly addressed intracavity fields and near‑field radiation, leaving open the key question of how two‑dimensionally confined polaritonic modes are converted into three‑dimensional radiative channels accessible in the far field.  In particular, the diffraction‑mediated outcoupling at cavity apertures and its spectral and angular fingerprints on the measurable emissivity, remain largely unexplored as it has neither been theoretically quantified nor experimentally probed yet.

In this work, we experimentally and theoretically investigate the far-field thermal emission from planar Si/SiO$_2$ cavities supporting the propagation of TEGMs along the SiO$_2$-vacuum interfaces (Fig.~\ref{fig:fig1}). 
By combining hemispherical emissivity measurements with fluctuational electrodynamics simulations, we identify the fingerprints of these confined modes and demonstrate that they can efficiently couple to the far field outside the cavity, significantly enhancing emissivity beyond that of flat surfaces. We show that SiO$_2$-covered cavities yield pronounced, spectrally narrow emissivity peaks and can enhance the emissivity by up to $200\%$ near the polaritonic resonance. Numerical simulations based on SCUFF-EM \cite{Guillemot2025,Reid2017,Rodriguez2013,Nguyen2017} reveal that this enhancement originates from the redistribution of electromagnetic energy density within the cavity and its diffraction through the aperture, which transforms the TEGMs into radiative channels over a finite angular range. These results establish a direct link between intracavity polaritonic modes and far‑field thermal emission, and position planar cavities as effective and scalable platforms for engineering high‑emissivity surfaces without resorting to complex nanostructuring. More broadly, this work highlights the great potential of simple cavity architectures for thermal photonics, passive cooling, infrared emission control, and thermophotovoltaic energy conversion.
\section*{Results}
\subsection*{Total Emissivity Measurements}
\begin{figure}[H]
\centering
\includegraphics[width=0.95\linewidth]{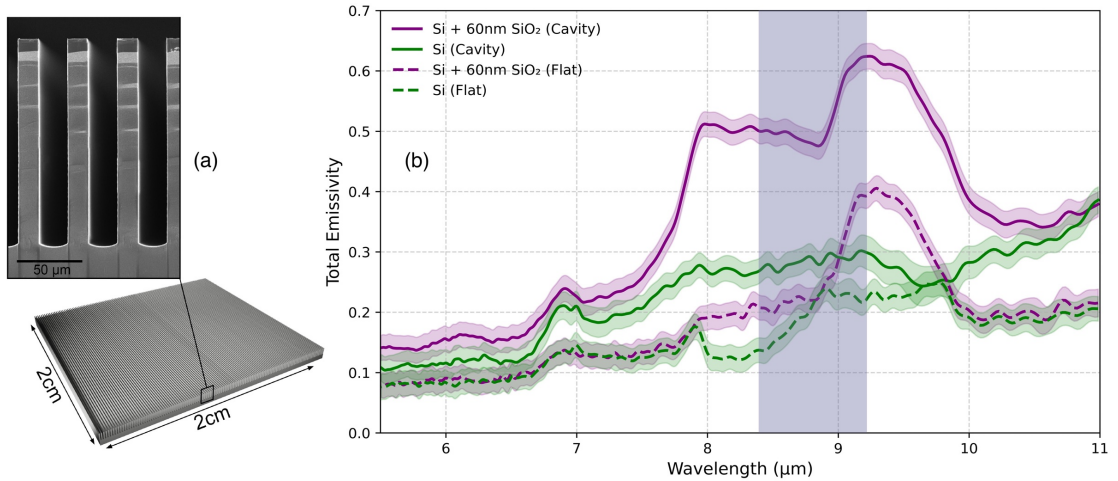}
\caption{\textbf{Total emissivity spectra of flat and cavity samples}. (a) Scanning electron microscope (SEM) image of the fabricated cavities, comprising 20-$\upmu$m-wide and 160-$\upmu$m-deep gaps separated by 20-$\upmu$m-thick Si walls. (b) Room-temperature total hemispherical emissivity measured for flat reference samples (dashed lines) and cavity samples (solid lines) made of bare Si (green) and Si covered with a 60-nm-thick thermally grown SiO$_2$ layer (purple). The shaded purple band marks the spectral range in which SPhPs exist for a SiO$_2$/Vacuum/SiO$_2$ double interface (see SM Fig.~S2), highlighting the resonance window associated with the cavity-induced emissivity enhancement. Bare Si cavities exhibit a broadband emissivity increase due to geometric effects, whereas the SiO$_2$-covered cavities exhibits a pronounced spectral peak consistent with the thermal emission mediated by TEGMs.}
\label{fig:fig4}
\end{figure}

We first investigate the experimental signature of TEGMs through infrared emissivity measurements using a Fourier-Transform InfraRed (FTIR) spectrometer. The samples consisted of two geometries and two materials: flat surfaces and planar cavities made up of bare silicon (Si) and Si covered with a thermally grown SiO$_2$ layer of 60~nm in thickness to activate SPhPs \cite{Tachikawa2022APL}. The cavity vacuum gaps, 20~$\upmu$m wide and 160~$\upmu$m deep, were separated by 20-$\upmu$m-thick Si walls, as illustrated in Fig.~\ref{fig:fig4}(a). The flat samples, with and without the SiO$_2$ layer, served as references to isolate cavity-induced effects in the presence and absence of SPhPs, respectively.

We performed total hemispherical emissivity measurements using an integrating sphere, which captures both the specular and diffuse components. This configuration ensures the collection of the spatially averaged emissivity over the $5–11.5~\upmu$m wavelength range. The specular emissivity measured with the incident beam normal to the cavity aperture did not reveal any detectable TEGM signature, as shown in Fig.~S1 of the Supplementary Material (SM). 

Figure~\ref{fig:fig4}(b) shows the total hemispherical emissivity spectra of flat (dashed) and cavity (solid) samples of bare Si (green) and Si covered with a SiO$_2$ nanolayer (purple). For bare Si, surface patterning into cavities yields a broadband emissivity enhancement across the full considered spectrum, without introducing pronounced spectral features. For the wavelength of 8.5~$\upmu$m, the emissivity increases from 0.15, for the flat sample, to 0.27, for the cavity sample, consistent with earlier studies attributing such broadband enhancement to the increased surface area and multiple internal reflections induced by surface structuring  \cite{Zhao2022}. For flat samples, the SiO$_2$ nanolayer introduces a pronounced emissivity peak at 9.5~$\upmu$m, associated with the lattice vibration and intrinsic SPhP response of SiO$_2$ \cite{Kovacevic2000,Volz2022}. This SPhP peak increases the emissivity from 0.40 (flat sample) to 0.65 (cavity sample), demonstrating that the SiO$_2$ nanolayer in Si cavities enhances the emissivity more effectively than in flat Si. Importantly, the cavity induces a second peak at 8.5$~\upmu$m, which raises emissivity from 0.15 (flat) to 0.50 (cavity). This additional peak only appears when both the cavity and the SiO$_2$ nanolayer are present, pointing to an origin rooted in the interplay between cavity confinement and SPhP propagation. This hypothesis is supported by the spectral window (purple band in Fig.~\ref{fig:fig4}(b)) allowing the existence and propagation of SPhPs along a SiO$_2$/vacuum/SiO$_2$ double interface (SM Fig.~S2) and spanning over the 8.5~$\upmu$m peak. The cavity thus selectively enhances emissivity within the SPhP spectral window, consistent with the emission mediated by TEGMs \cite{Volz2022}. The propagation of TEGMs and the emissivity enhancement are not expected to be significantly affected by roughness, because the Bosch process used to fabricate the cavities (see Methods) typically yields sidewall roughness ($100–400$~nm RMS) \cite{Hayashi19,Fu18} much smaller than the SPhP wavelength ($8–10~\upmu$m) \cite{Volz2022}. This large dimensional mismatch justifies the smooth-wall modeling used in the simulations below.

\subsection*{Spatial Distribution of the Cavity-Induced Energy Density}
\begin{figure}[H]
\centering
\includegraphics[width=\linewidth]{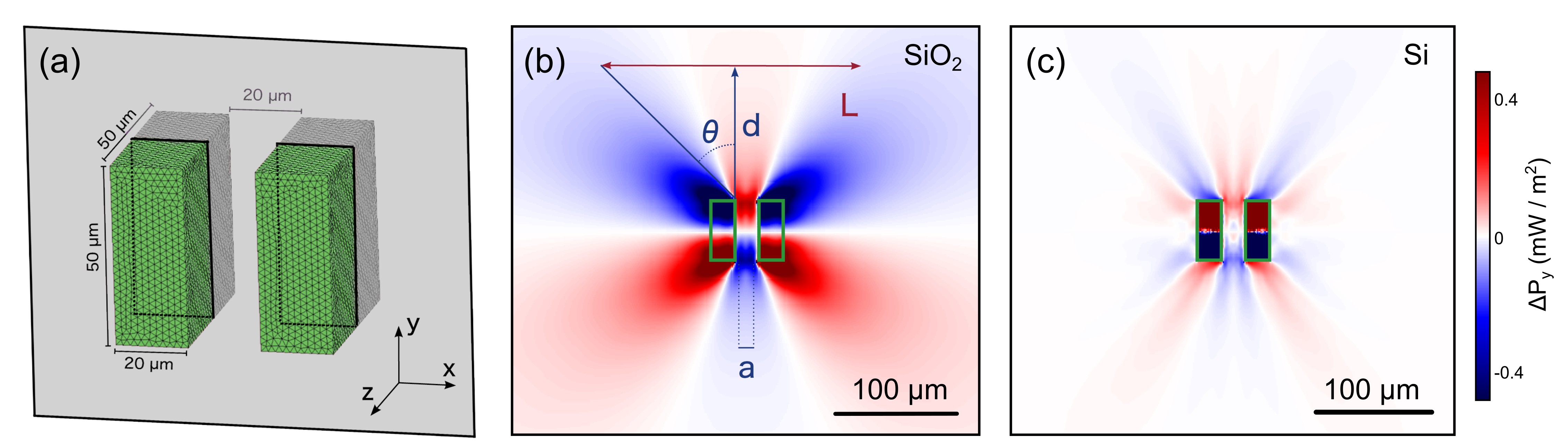}
\caption{\textbf{Spatial distribution of the cavity-induced Poynting vector}.(a) Mesh used in the SCUFF-EM simulations for the cavity walls. The calculations were performed on a plane located at $z = 25~\upmu$m (grey square), which corresponds to the midpoint of the walls' width. (b) Frequency-integrated map of the cavity-induced Poynting vector $\Delta P_y$ for a cavity with SiO$_2$ walls outlined by the green rectangles. The annotations identify the aperture size $a$, observation distance from the aperture $d$, diffraction angle $\theta$, and computational window $L$. (c) $\Delta P_y$ map obtained for a cavity with Si walls under identical conditions. Positive and negative values of $\Delta P_y$ represent the local enhancement and suppression of the radiative flux relative to the sum of isolated wall contributions. The SiO$_2$ cavity concentrates energy along the cavity channel and aperture, enabling waveguide-like transport toward the far field, whereas the bare Si cavity displays a response dominated by edge effects and internal redistribution rather than guided emission.}
\label{fig:fig2}
\end{figure}

To elucidate the physical origin of the observed emissivity enhancement and understand the energy transport mechanism out of the cavity, we performed fluctuational electrodynamics simulations using the solver SCUFF-EM \cite{Guillemot2025,Reid2017,Rodriguez2013,Nguyen2017}. These numerical calculations provide spatially and spectrally resolved maps of the Poynting vector in and out the cavity. Rather than analyzing raw fluxes, we define the cavity‑induced Poynting vector $\Delta P$ as the difference between the total flux radiated by the full cavity and the superposition of the fluxes radiated independently by each wall (see Methods). Subtracting single-wall contributions isolates the interface and coupling effects that are intrinsic to the cavity geometry. The simulated geometry consists of 20-$\upmu$m-thick and 50-$\upmu$m-tall SiO$_2$ (or Si) walls separated by a 20~$\upmu$m gap, as shown in Fig.~\ref{fig:fig2}(a). Since SPhPs primarily propagate along the SiO$_2$/vacuum interfaces and evanescently decay over a cross-plane distance of $\sim$50~nm \cite{Volz2022}, much shorter than the 20~$\upmu$m wall thickness, the experimental Si walls covered with a 60-nm-thick SiO$_2$ layer can be approximated as bulk SiO$_2$ walls. This model thus captures the essential physics of the fabricated cavities while remaining computationally tractable. The walls' temperature was set to $T_{\mathrm{obj}} = 301$~K and the environment to $T_{\mathrm{env}} = 300$~K to match experimental conditions.

Figure~\ref{fig:fig2}(b) shows the frequency-integrated map of $\Delta P_y$, the $y-$component of $\Delta P$ accounting for the electromagnetic flux normal to the aperture of the cavity with SiO$_2$ walls. For $y > 0$, positive (negative) $\Delta P_y$ values denote local flux enhancement (reduction) induced by the cavity. This interpretation naturally reverses for $y<0$ due to the system's symmetry. The SiO$_2$ cavity exhibits a waveguide-like response, reducing the energy density in all directions except within the cavity and along its aperture. This spatial distribution indicates that the structure effectively extracts energy from its surroundings and channels it through the gap along the axial direction. Therefore, unlike a conventional resonant cavity, the SiO$_2$ structure functions as a waveguide that efficiently directs energy toward the far field. In addition, the channeled cavity-induced flux does not emerge only at normal incidence ($\theta = 0°$), as it undergoes diffraction at the aperture edges, spreading the emitted radiation over a finite angular range. This angular redistribution of energy naturally transforms laterally guided modes into radiative channels accessible in hemispherical measurements. By contrast, in the case of Si walls shown in Fig.~\ref{fig:fig2}(c), the cavity strongly enhances the flux inside the walls and redistributes it around their outer edges, indicating that the emission enhancement is dominated by edge effects. Here, the cavity acts as a geometric resonator that boosts internal reflections rather than guiding energy preferentially through the aperture.

\section*{Discussion}
\begin{figure}[H]
\centering
\includegraphics[width=0.95\linewidth]{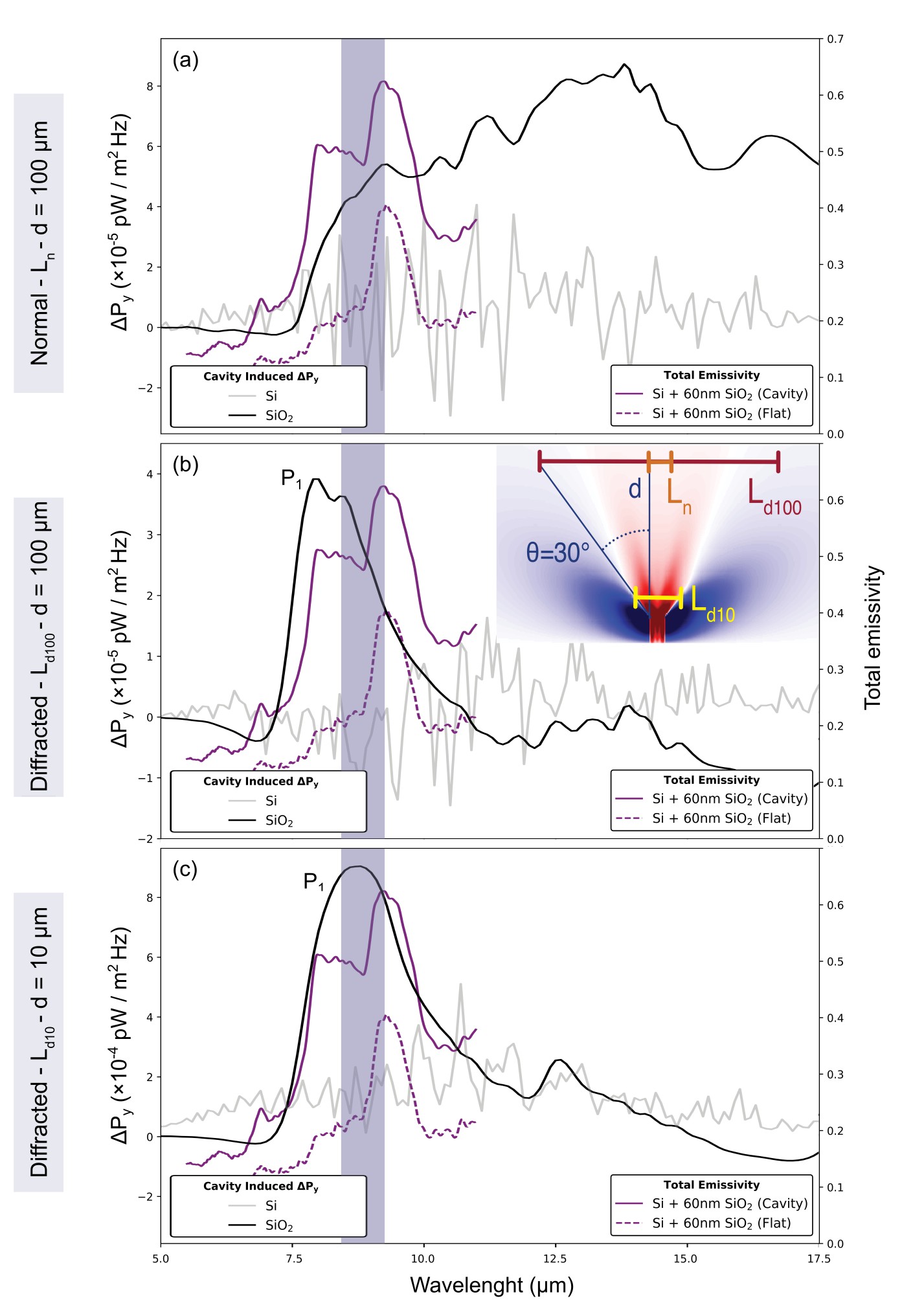}
\caption{\textbf{Spectrally resolved cavity-induced Poynting vector and measured emissivity spectra.} Left axis: Poynting vector $\Delta P_y$ calculated for cavities with SiO$_2$ (black) and Si (grey) walls. The spectra are averaged over the spatial window (a) $L_n$ ($\theta = 0^\circ$) at $d=100~\upmu$m, (b) $L_{d100}$ ($\theta = 30^\circ$) at $d=100~\upmu$m, and (c) $L_{d10}$ ($\theta = 30^\circ$) at $d=10~\upmu$m. Right axis: Experimental emissivity spectra reproduced from Fig.~\ref{fig:fig4} for flat and cavity samples covered with a 60-nm-thick SiO$_2$ layer. The shaded purple region indicates the SPhP spectral range for a SiO$_2$/vacuum/SiO$_2$ double interface (see SM Fig.~S2). The Si cavity shows no distinct spectral signature, in agreement with the broadband emissivity enhancement observed experimentally. By contrast, the SiO$_2$ cavity exhibits a pronounced diffracted peak near $8.8~\upmu$m at short distance that evolves into a doublet in the far field, matching the measured emissivity enhancement near $8.5~\upmu$m. These results thus identify the far-field signature of TEGMs as the diffracted manifestation of polariton-cavity hybridization.}
\label{fig:fig3}
\end{figure}

To identify the spectral signature of the directed flux, we computed the spectrally resolved Poynting vector $\Delta P_y$ and averaged it over a spatial window $L = a + 2d \tan\theta$ (Fig.~\ref{fig:fig2}b) at a distance $d$ from the aperture of width $a$ and for a diffraction angle $\theta$ obeying the grating equation: $a\sin\theta = \lambda$ \cite{BornWolf1999}.
The angle $\theta$ thus sets the limits of the main diffraction lobe outside the cavity. The averaged flux is shown in Fig.~\ref{fig:fig3} for the distances $d=100~\upmu$m and $d=10~\upmu$m, and two representative diffraction angles: $\theta=0^\circ$ corresponding to the normal component, and $\theta=30^\circ$ representing the diffraction angle for the Wien's wavelength $\lambda=9.66~\upmu$m at 300~K.

For both the normal and diffracted components of the $\Delta P_y$ spectra, the Si cavity does not exhibit distinct spectral features across the considered wavelength range, in agreement with the broadband behavior observed experimentally (see Fig.~\ref{fig:fig4}). This featureless spectrum confirms that purely geometric cavity effects, in the absence of SPhPs, do not produce narrow-band emissivity enhancements. In stark contrast, the SiO$_2$ cavity displays a strong sensitivity of energy density to both angle and distance. The normal component of $\Delta P_y$ remains featureless within the SPhP existence range, indicating no clear interplay between SPhPs and the cavity. The diffracted component, however, reveals pronounced TEGM signatures. For $d=10~\upmu$m, we observe a single peak $P_1$ at $8.8~\upmu$m that lies near the center of the SPhP spectrum. The emergence of $P_1$, showing enhanced emissivity attributable to the SiO$_2$ cavity geometry, demonstrates its SPhP‑related origin and constitutes a clear near‑aperture TEGM signature. As the observation plane moves to $d=100~\upmu$m in the far field, $P_1$ evolves into a doublet with maxima at $7.9~\upmu$m and $8.4~\upmu$m. This splitting closely matches the experimental emissivity peak near $8.5~\upmu$m. The spectral correspondence between the simulated $\Delta P_y$ and measured emissivity confirms that the far‑field $P_1$ peak represents the diffracted manifestation of TEGMs. Moreover, the fact that TEGM signatures arise exclusively in the diffracted component explains their absence in specular emissivity measurements performed at normal incidence.

Taken together, these results demonstrate that the emissivity enhancement in SiO$_2$-covered cavities originates from thermally excited guided modes resulting from the polariton-cavity hybridization. The cavity transforms two‑dimensionally confined polaritonic modes into controllable three‑dimensional thermal emission through diffraction at its aperture, thereby establishing TEGMs as an efficient channel for far-field thermal radiation. This mechanism offers a simple and scalable approach to tailor the spectral and angular properties of thermal emission using planar cavities without resorting to complex nanofabrication.

\section*{Methods}
\subsection*{Numerical simulations}
Numerical calculations were performed under the framework of fluctuational electrodynamics using the open-source solver SCUFF-EM \cite{Rodriguez2013,Nguyen2017}. This solver implements the Fluctuation-Dissipation Theorem (FDT) through a surface-integral formulation of Maxwell's equations that discretizes only material boundaries via the Boundary Element Method (BEM). In this formalism, thermally fluctuating volumetric currents are reformulated as equivalent surface currents, enabling the efficient evaluation of radiative heat transfer and field distributions in structures of arbitrary geometry.

SCUFF‑EM provides frequency- and space-resolved observables, including the Poynting vector and the Maxwell stress tensor. The temporally averaged Poynting vector used in this work is implemented in SCUFF-EM as follows:
\begin{equation}
\langle P_y(\mathbf{r},u)\rangle = \frac{\hbar\omega_0}{2}\sum_b \Delta \hat{\Theta}_b(u)\,\Phi_{\mathbf{r}'\rightarrow \mathbf{r}}(u),
\label{eq:poynting_scuff}
\end{equation}
where $u=\omega/\omega_0$ is a normalized frequency with $\omega_0 = 3\times 10^{14}$~rad/s; $\Delta \hat{\Theta}_b(u)=\hat{\Theta}_b(u)-\hat{\Theta}_{\mathrm{env}}(u)$ is the difference between the Bose-Einstein distribution of the source body $b$ at $\mathbf{r}'$ and that of the environment, and $\Phi_{\mathbf{r}'\rightarrow \mathbf{r}}(u)$ is the temperature-independent generalized flux describing the contribution of current fluctuations in $\mathbf{r}'$ to the energy density at $\mathbf{r}$.

\begin{figure}[H]
\centering
\includegraphics[width=1\linewidth]{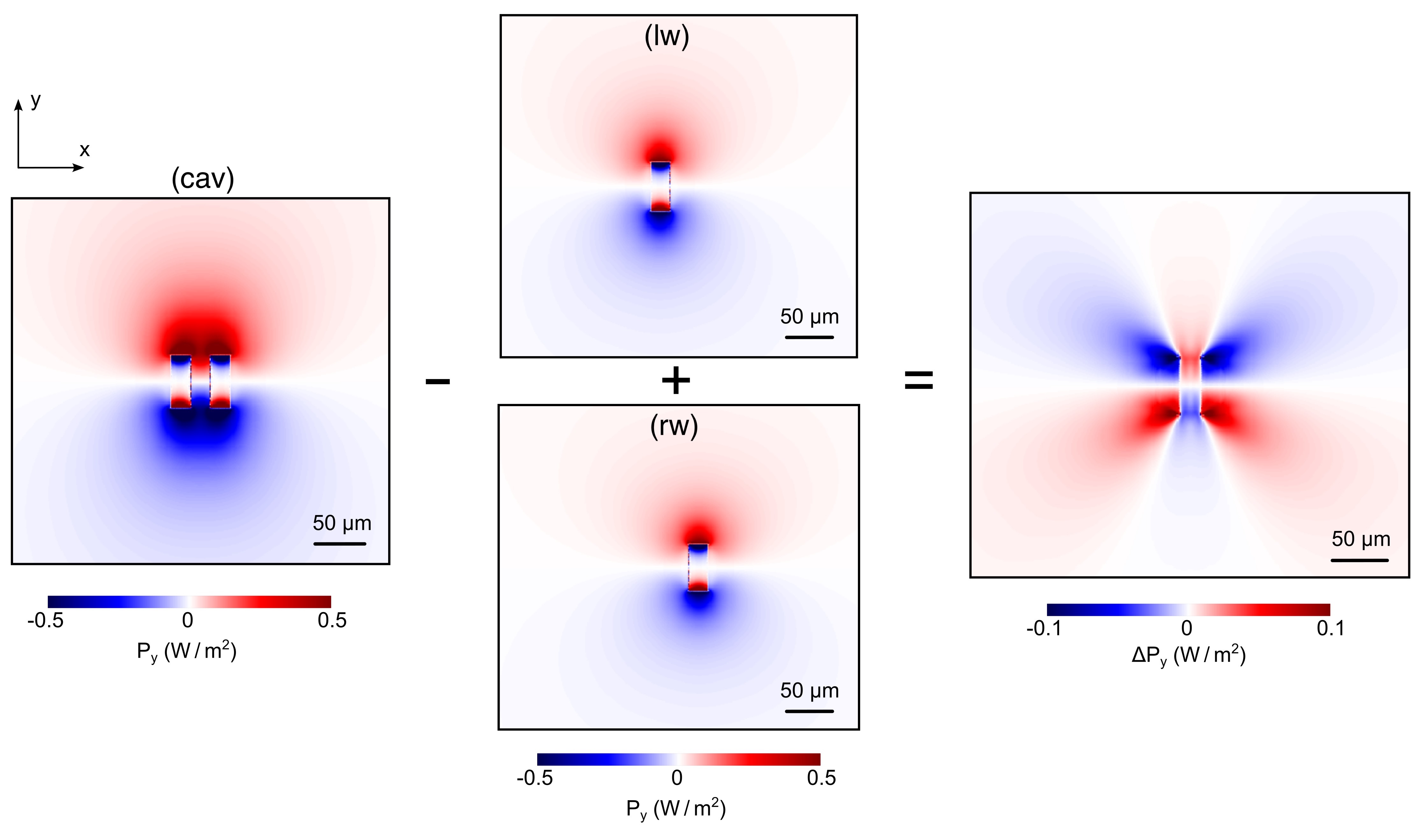}
\caption{\textbf{Decomposition of the cavity-induced Poynting vector into full-cavity and single-wall contributions}. Maps of the cavity-induced Poynting vector $P_y$ obtained from fluctuational electrodynamics simulations. The panel labeled "cav" shows the result for the complete two-wall cavity, whereas "lw" and "rw" display the corresponding maps obtained for the isolated left and right walls, respectively. The subtraction of the single-wall contributions isolates the hybridization effect intrinsic to the cavity geometry and removes the flux arising from individual wall emission. These maps can be evaluated at a single frequency or integrated over the entire spectral range, enabling direct visualization of the electromagnetic energy redistribution induced by the cavity.}
\label{fig:fig5}
\end{figure}

To isolate the effect of the cavity geometry, we define the following cavity-induced Poynting vector as the difference between the total flux of the full cavity and the fluxes radiated independently by each wall:
\begin{equation}
\Delta P_{y} = P_{y}(\mathrm{cav}) - P_{y}(\mathrm{lw}) - P_{y}(\mathrm{rw}),
\label{eq:deltaP}
\end{equation}
where "cav" denotes the complete two-wall cavity, and "lw" and "rw" refer to the isolated left and right walls, respectively. This subtraction removes single-wall contributions and highlights the mode hybridization intrinsic to the cavity configuration, as shown in Fig.~\ref{fig:fig5}.

\subsection*{Sample preparation}
Samples were fabricated from a high‑resistivity undoped silicon wafer (resistivity$>10~$k$\Omega\cdot$cm, thickness of 380~$\upmu$m). Arrays of rectangular cavities with depth $\sim$160~$\upmu$m and width $\sim$20~$\upmu$m were etched using Deep Reactive Ion Etching (DRIE) with the Bosch process. We thus obtained 20-$\upmu$m‑wide gaps separated by 20-$\upmu$m‑thick Si walls, as illustrated in Fig.~\ref{fig:fig4}. To activate SPhPs, the cavity walls and top surfaces were subsequently covered with a 60-nm‑thick SiO$_2$ layer grown by thermal oxidation. The fabrication process is schematically summarized in SM Fig.~S3.

\subsection*{Optical measurements}
The spectral emissivity $\varepsilon$ of the samples was determined from the reflectance ($R$) and transmittance ($T$) measurements performed with a Fourier-Transform InfraRed (FTIR) spectrometer (JASCO FT/IR-8X, spectral resolution of 4~cm$^{-1}$) coupled to an integrating sphere (PIKE Technologies, gold-coated interior, inner diameter of 7.62~cm, incident beam angle of 12$^\circ$). The integrating sphere collects both the specular and diffuse components of the reflected energy, allowing the measurement of the total hemispherical emissivity over the wavelength range $5-11.5~\upmu$m. The infrared signal was detected at room temperature using a liquid-nitrogen‑cooled MCT (mercury cadmium telluride) detector. Considering that our system was in thermal equilibrium, we used Kirchhoff's law of thermal radiation (emissivity = absorptivity) and the principle of energy conservation to determine the spectrum of room-temperature hemispherical emissivity through the relation $\varepsilon(\lambda)=1-R(\lambda)-T(\lambda)$. 

For each sample, ten spectra were recorded with random in‑plane orientations to assess reproducibility and average out possible anisotropies (see SM Fig.~S4). We estimated the experimental uncertainty from the standard deviation over repeated measurements on two independently fabricated samples with nominally identical properties. We added a $\pm$2\% uncertainty to the standard deviation to account for fabrication variability (see SM Fig.~S5 for details)


\section*{Acknowledgements}
We acknowledge the valuable support of Byunggi Kim, Yu-Bin Chen, Hao-Yu Kang, and Elissa Akiki for carrying out some preliminary measurements. This work was supported by the CREST JST (Grant No. JPMJCR19I1) and KAKENHI JSPS (Grant Nos. 21H04635 and JP20J13729) projects.

\section*{Author contributions}
The sample fabrication and experiments were performed and analyzed by M. C., L. J., and G. H. under the supervision of M. N., J. W., and S. V. Modeling was carried out by M. C. in collaboration with R. A. and J. O.-M. The manuscript was written by M. C. and J. O.-M. and revised by all authors. All authors discussed the experimental and theoretical results. 

\section*{Competing interests}
The authors declare no competing interests.
\end{document}